\documentclass[11pt]{llncs}

\usepackage{latexsym}
\usepackage{txfonts}
\usepackage{graphicx}
\usepackage{ascmac}
\usepackage{color}

\usepackage{hangcaption}
\usepackage{slashbox}
\usepackage{comment}
\usepackage{wrapfig}
\usepackage{multirow,eepic}

\renewcommand{\date}{\today}

\makeatletter
\spn@wtheorem{observation}{Observation}{\itshape}{\rmfamily}
\makeatother

\newcommand{\nop}[1]{}

\makeatletter
\def\eqnarray{%
\stepcounter{equation}%
\let\@currentlabel=\theequation
\global\@eqnswtrue
\global\@eqcnt\z@
\tabskip\@centering
\let\\=\@eqncr
$$\halign to \displaywidth\bgroup\@eqnsel\hskip\@centering
$\displaystyle\tabskip\z@{##}$&\global\@eqcnt\@ne
\hfil$\displaystyle{{}##{}}$\hfil
&\global\@eqcnt\tw@$\displaystyle\tabskip\z@{##}$\hfil
\tabskip\@centering&\llap{##}\tabskip\z@\cr}
\makeatother

\makeatletter
\def\senbun#1(#2)#3({\@senbun(#2)(}
\def\@senbun(#1,#2)(#3,#4){%
   \@tempdima#1\p@ \advance\@tempdima#3\p@
   \divide\@tempdima\tw@
   \@tempdimb#2\p@ \advance\@tempdimb#4\p@
   \divide\@tempdimb\tw@
   \edef\@senbun@temp{\noexpand\qbezier(#1,#2)%
      (\strip@pt\@tempdima,\strip@pt\@tempdimb)(#3,#4)}%
   \@senbun@temp}
\makeatother

\newcounter{enum2}
\newenvironment{enumerate2}{%
   \begin{list}%
   {%
      \arabic{enum2}.\ \,
   }%
   {%
      \usecounter{enum2}
      \setlength{\topsep}{0ex}
      \setlength{\itemindent}{0em}
      \setlength{\leftmargin}{3em}
      \setlength{\rightmargin}{0em}
      \setlength{\labelsep}{0em}
      \setlength{\labelwidth}{3em}
      \setlength{\itemsep}{0.0em}
      \setlength{\parsep}{0em}
      \setlength{\listparindent}{0em}
   }
}{%
   \end{list}%
}

\graphicspath{{./figure/}{../fun12_submit/figure/}{../algorithms/figure/}}

\title{Solving Tantrix via Integer Programming}

\author{%
   Fumika Kino\inst{1} \and 
   Yushi Uno\inst{2}
}
\institute{
    Mitsubishi Electric Information Network Corp., 
    8-1-1 Tsukaguchi-Honmachi, Amagasaki 661-8611, Japan. 
    \email{tanukinoko0049@gmail.com}
\and
    Graduate School of Science, Osaka Prefecture University, 
    1-1 Gakuen-cho, Naka-ku, Sakai 599-8531, Japan. 
    \email{uno@mi.s.osakafu-u.ac.jp}
}

\begin{document}
\maketitle

\begin{abstract}
Tantrix 
is a puzzle to make a loop 
by connecting lines drawn on hexagonal tiles, 
and the objective of this research is to solve it by a computer. 
For this purpose, we give a problem setting of solving Tantrix 
as arranging tiles in an appropriate shape and making a loop at the same time 
within a given hexagonal lattice board. 
We then formulate it as an integer program 
by expressing the rules of Tantrix as its constraints, 
and solve it by a mathematical programming solver to have a solution. 
As a result, we establish a formulation that solves Tantrix 
of moderate sizes, 
and even when the solutions are invalid only by elementary constraints, 
we achieved it by introducing additional constraints 
and an artificial objective function 
to avoid flaws in invalid solutions. 
By this approach 
we are successful in solving Tantrix of size up to 50. 
\end{abstract}

\renewcommand{\thefootnote}{\fnsymbol{footnote}}
\setcounter{footnote}{4}

\section{Introduction}
\label{introduction}

Games and puzzles are entertainments invented for human beings, 
and solving puzzles or playing games are lots of fun for everybody. 
Such puzzles and games are often logical enough, 
and they have long been attracted interests 
of mathematicians and computer scientists 
not only for the pleasure but for their research \cite{G05}. 
Those may include Nim, Hex, Peg Solitaire, Tetris, Geography, Sudoku, 
Rubik's Cube, Chess, Othello, Go, and so on 
\cite{sudoku,G05,LS80,rubik,R82,S50}. 

There are a lot of directions and objectives 
when puzzles and games are treated as research topics 
\cite{ANW07,D01,G05,HD09}. 
Some of those are 
investigating their mathematical structures \cite{sudoku,L03}, 
computational complexities \cite{DDUUU10,LS80}, 
winning strategies \cite{ANW07,G05}, and so on. 
As computers evolve and expand their applicational utilization, 
they are rapidly incorporated into these research areas 
\cite{sudoku,rubik,R82,S50}. 
Typical example is to develop a computer program 
that can solve puzzles faster than humans 
or can beat humans in playing games. 

In this paper, we focus on a puzzle called 
Tantrix\footnote{
Tantrix$\mbox{}^{\scriptsize\textregistered}$ 
is a registered trademark of Colour of Strategy Ltd. in New Zealand, 
and of TANTRIX JAPAN in Japan, respectively, 
under the license of M.~McManaway, the inventor. 
}
that makes a loop by connecting lines drawn on hexagonal tiles 
\cite{Tantrix-JP,Tantrix-NZ,Tantrix-UK}, 
and the objective of this research is to solve it by a computer\footnote{
Solutions obtained by computers are not authorized as official records.
}. 
More precisely, we first give a problem setting of solving Tantrix 
as arranging tiles in an appropriate shape and making a loop at the same time 
within a given hexagonal lattice board. 
We then formulate it as an integer program (IP) \cite{P03,W99} 
by expressing the rules or properties that solutions of Tantrix satisfy 
(necessary conditions) as its constraints, 
and we attempt to obtain a solution by solving that IP 
using a commercial mathematical programming solver. 
However, since we may not have valid solutions 
only by elementary constraints, 
we develop some new additional constraints 
and introduce an artificial objective function 
in order to derive valid solutions. 
We show the current best solution of Tantrix 
obtained by the proposed approach. 
Our approach of using IP formulation to solving puzzles seems 
novel and entertaining, 
and to the best of our knowledge, 
this is one of few cases that IP meets solving puzzles successfully. 

In Section~\ref{tantrix}, 
we introduce a puzzle called Tantrix and explain its rules. 
Next in Section~\ref{preliminary}, 
we give a problem setting for solving Tantrix by a computer 
and terminology for IP formulations. 
Then in Section~\ref{formulation}, 
we describe an elementary integer programming formulations of Tantrix 
and show some computational results. 
In Section~\ref{improvement}, we develop novel ideas 
to obtain valid solutions and show our current best result. 
Finally Section~\ref{conclusion} presents some future work 
and concludes the paper.

\section{Tantrix}
\label{tantrix}

Tantrix is a puzzle originally invented in 1988 
by Mike McManaway of New Zealand \cite{Tantrix-JP}. 
Several variants of commercial Tantrix products have been sold so far, 
and among those a {\rm solitaire} version is named ``Tantrix Discovery'' 
\cite{Tantrix-JP,Tantrix-NZ,Tantrix-UK,Tantrix-TPguide}. 
Throughout this paper, we fucus only on this solitaire version, 
Tantrix Discovery, 
and we simply call it ``Tantrix''. 

\begin{figure}[htb]
 \begin{center}
  \includegraphics[height=25.5mm]{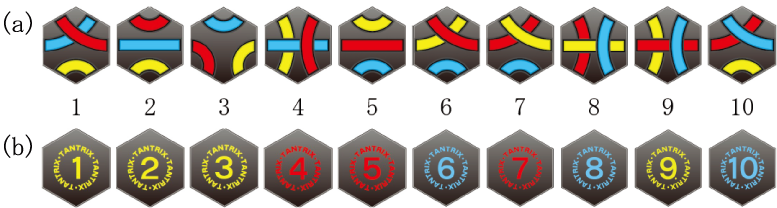}
 \end{center}
  \caption{(a) Tops of 10 sorts of tiles (with their orientations 1), 
and (b) their corresponding backs.} 
\label{tile_surfaces}
\end{figure}

Tantrix is played with 10 sorts of hexagonal tiles of the same size. 
A {\it tile} has two surfaces, 
which we call a {\it top} ({\it surface}) and a {\it back} ({\it surface}). 
On a top surface three lines are drawn in red, blue and yellow 
(Fig.~\ref{tile_surfaces} (a)), 
and on a back surface one of the numbers from 1 to 10 is drawn 
in either one of the three colors 
(Fig.~\ref{tile_surfaces} (b)). 
The 10 patterns of lines drawn on tops are all different from each other. 

\begin{figure}[htb]
\begin{minipage}{.02\linewidth}
\mbox{}
\end{minipage}
\begin{minipage}{.580\linewidth}
 \begin{center}
  \includegraphics[height=28.5mm]{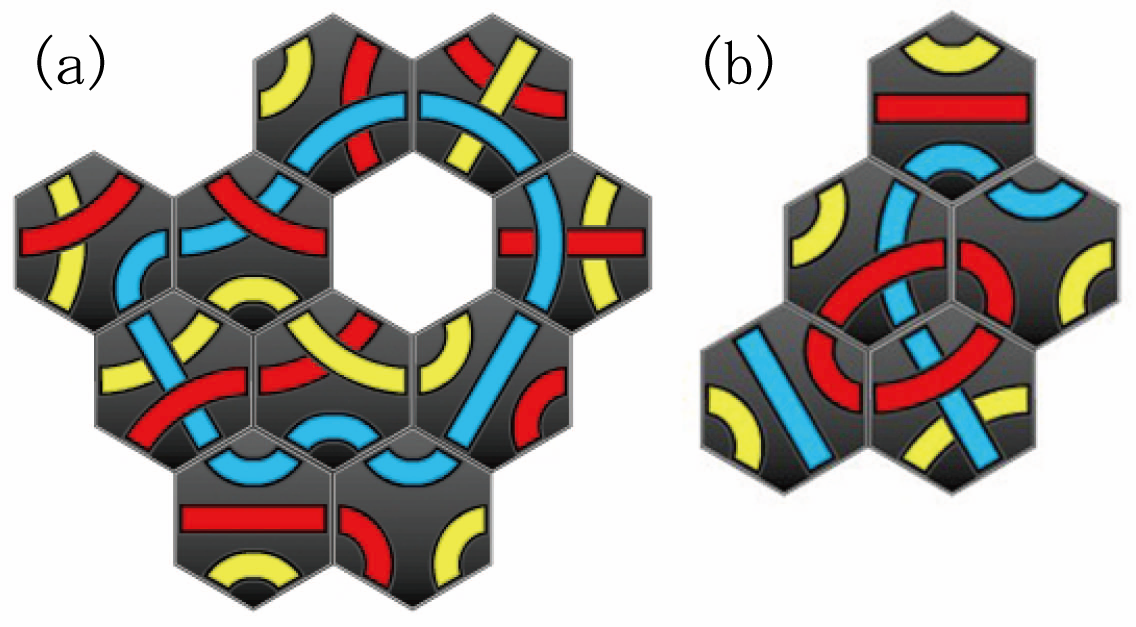}
 \end{center}
  \caption{Examples of ``uncleared'' arrangements 
although they satisfy rules from 1 to 5: 
(a) there is a hole, and 
(b) a line of the designated color (red) is not a part of a loop.}
 \label{out_of_rules}
\end{minipage}
\begin{minipage}{.03\linewidth}
\mbox{}
\end{minipage}
\begin{minipage}{.325\linewidth}
 \begin{center}
  \vspace*{0.2cm}
  \includegraphics[height=23.0mm]{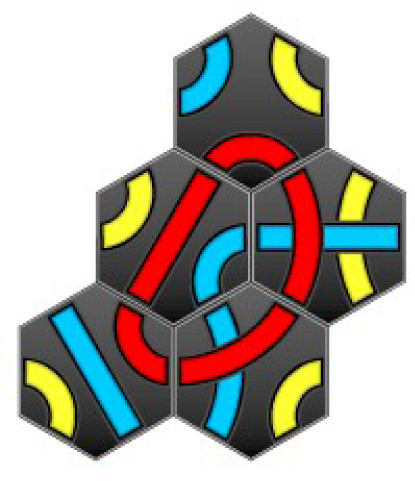}
\vspace*{-0.3cm}
 \end{center}
  \vspace*{0.3cm}
  \caption{An example of a ``cleared'' arrangement; 
a Tantrix solution of challenge number 5.} 
 \label{challenge5}
\vspace*{-0.3cm}
\end{minipage}
\begin{minipage}{.02\linewidth}
\mbox{}
\end{minipage}
\end{figure}

Tantrix is played by putting tiles in the form of hexagonal lattice 
and the goal is to make a loop in one designated color 
according to the following rules: 
\begin{enumerate2}
\item 
determine the number of tiles for the challenge ({\it challenge number}), 
which is greater than 2, 
\item 
prepare tiles as many as the challenge number 
by starting from a tile with number 1 
and consecutively up to that number 
(in case it is more than 10, start from 1 again), 
\item 
the {\it designated color} of the challenge number 
is the one in which its lowest digit 
is written (on the back of the tile), 
\item 
connect all the lines of the designated color (drawn on prepared tiles) 
so that they form a single loop, 
\item 
connect the lines of the other colors 
so that their touching colors match. 
\end{enumerate2}
Here arrangements of tiles in which there is a {\it hole} 
(places without tiles surrounded by 6 tiles or more) 
(Fig.~\ref{out_of_rules} (a)), 
or any one of the lines of the designated color is not a part of a loop 
(Fig.~\ref{out_of_rules} (b)) 
are not allowed. 
When one completes making a loop according to the above rules, 
we say that she/he {\it cleared} that challenge number of Tantrix. 
In addition to the above rules, 
to avoid that a solution becomes a repetition of a specified 
pattern, its arrangement is required to be, intuitively speaking, 
`round' (not flattend)\footnote{
To be authorized as an official record, 
an arrangement must satisfy the following condition: 
in the arrangement let one of the three directions (axises) 
that has the most number $x$ of tiles be A, 
and the other two directions that cross A be B and C. 
Then there must be more tiles than 30\% of $x$ in more rows 
than 75\% of those in directions B and C, respectively. 
This condition is set to exclude the so-called ``4-tiles' equation'' 
discovered by C.~Fraser of England. 
This information was once posted in \cite{Tantrix-JP}, however, 
we can no longer find it on Jan. 1, 2012. 
See Appendix~\ref{4tiles_equation} for details. 
}. 
(This leads our board setting explained later.)

Suppose we try Tantrix of challenge number 5. 
The designated color of that number is red (Fig.~\ref{tile_surfaces} (b)). 
In Fig.~\ref{challenge5}, a single loop of the designated color red is made 
by using all the red lines drawn on five tiles that have numbers from 1 to 5, 
and the colors of touching lines of the other colors also match 
(in this case blue only) at the same time. 
In addition there is no hole in this arrangement of tiles. 
Therefore, we say that we cleared Tantrix of challenge number 5. 

A {\it solution} (of challenge number $n$) 
is an arrangement how $n$ tiles are placed, 
and a {\it Tantrix solution} is the one 
that satisfies all the above conditions (Fig.~\ref{challenge5}). 
A {\it shape} of a solution (or a solution shape) is its boundary 
formed by $n$ hexagonal tiles.

\section{A Problem Setting and Terminology for Formulations}
\label{preliminary}

Our approach to solving Tantrix is to formulate it 
as an integer program 
and to solve it by mathematical programming solver. 
In this section, we give a problem setting for this purpose 
and some terminology for IP formulations.

\subsection{A Problem Setting for Solving Tantrix by a Computer}

\begin{figure}[htb]
 \begin{center}
  \includegraphics[height=26mm]{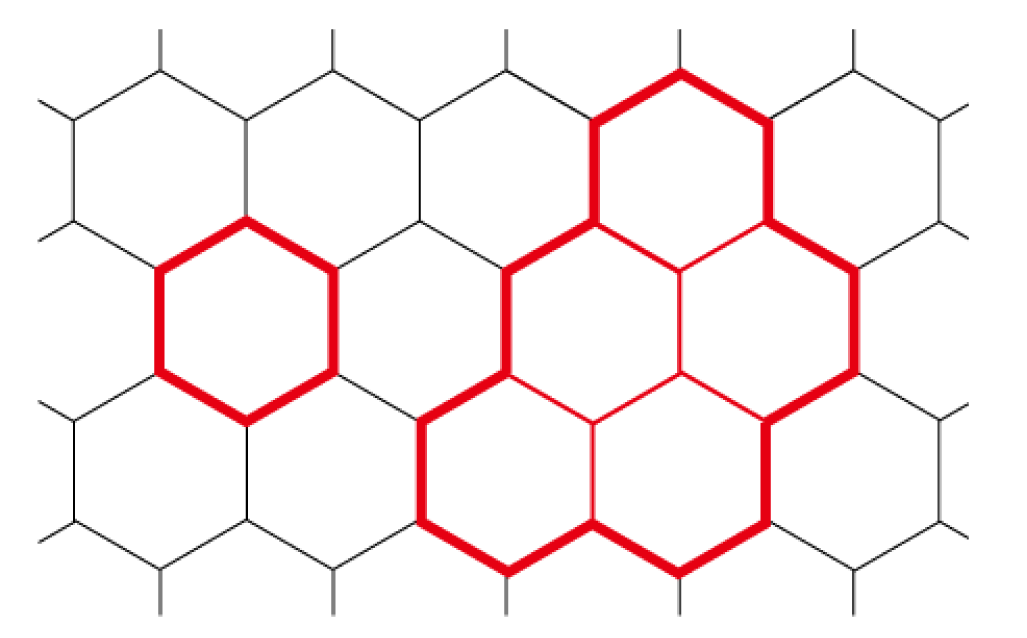}
 \end{center}
  \caption{A place (left) and an example of a board (of size 5) (right) 
on a hexagonal lattice plane.} 
 \label{board}
\end{figure}

Humans may play Tantrix on the table, 
on the floor or at any other places they like. 
For computers, however, we have to prepare artificially and appropriately 
a space where solutions are made. 
Consider an infinite hexagonal lattice plane (Fig.~\ref{board}), 
where the size of each hexagon is as the same as a single tile. 
When Tantrix is played by a computer, 
a tile is placed to fit on each hexagon, which we call a {\it place} 
(Fig.~\ref{board} (left)), 
and we call a collection of multiple places 
where a solution is supposed to be made a {\it board} 
(Fig.~\ref{board} (right)). 
The {\it size} of a board is the number of hexagons that constitute it. 

When a human solves Tantrix, one may connect the lines 
of the designated color to make a single loop, 
while connecting lines of the other colors simultaneously, 
as well as arranging tiles so that they do not create a hole 
and the shape become round at the same time. 
To realize these situations for a computer, 
especially to guarantee the freedom of solution shapes, 
we prepare a finite but sufficiently larger size of a board 
than the challenge number. 
That is, we describe the problem to be solved as follows: 

\begin{quote}
{\sc TANTRIX (Free\_Board TANTRIX)}\\
Input: a challenge number $n$ and a board of size $m$ $(>n)$, \\
Output: a Tantrix solution on (within) the input board. 
\end{quote}

\begin{figure}[htb]
\centering
\scalebox{0.63}{\includegraphics{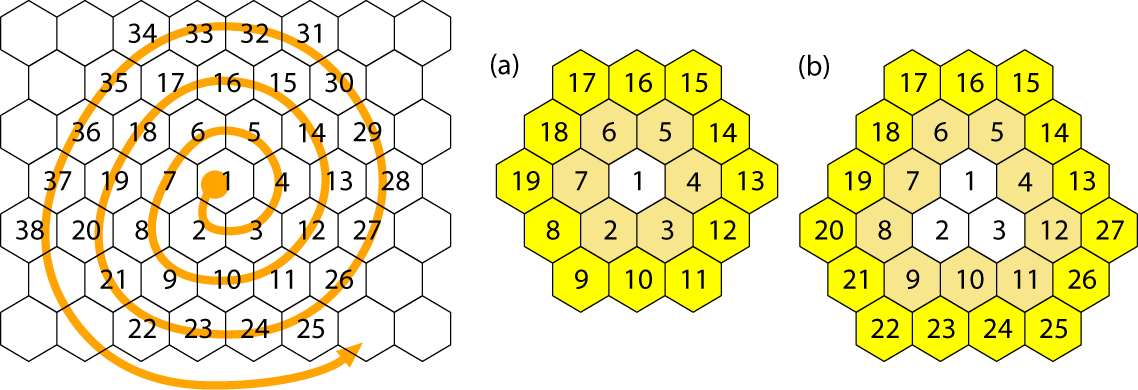}}
\caption{A numbering of places on infinite hexagonal plane, 
and boards of sizes (a) 19 of type~A and (b) 27 of type~B 
with their place numbers.}
\label{spiral_board}
\end{figure}

Here, since solutions are required to be arranged in round shapes, 
we give boards on which they are made in the folowing manner: 
number places from 1 in a spiral way starting from a single place 
(Fig.~\ref{spiral_board}), 
and pick consecutively numbered places (from 1) 
so that they become some symmetric shapes 
as shown in Fig.~\ref{spiral_board} (a) and (b) 
(we call them, for convenience, types~A and B, respectively). 
Therefore, we can prepare boards of sizes $7, 19, 37,\ldots ,$
and $3, 12, 27, 48,\ldots ,$ of types~A and B, respectively. 
When we try Tantrix of challenge number 20, for example, 
we prepare a board of size 27 of type~B or size 37 of type~A, and so on.

\subsection{Terminology and Definitions for IP Formulations}

\begin{figure}[htb]
\begin{minipage}{.02\linewidth}
\mbox{}
\end{minipage}
\begin{minipage}{.510\linewidth}
 \begin{center}
  \includegraphics[height=14mm]{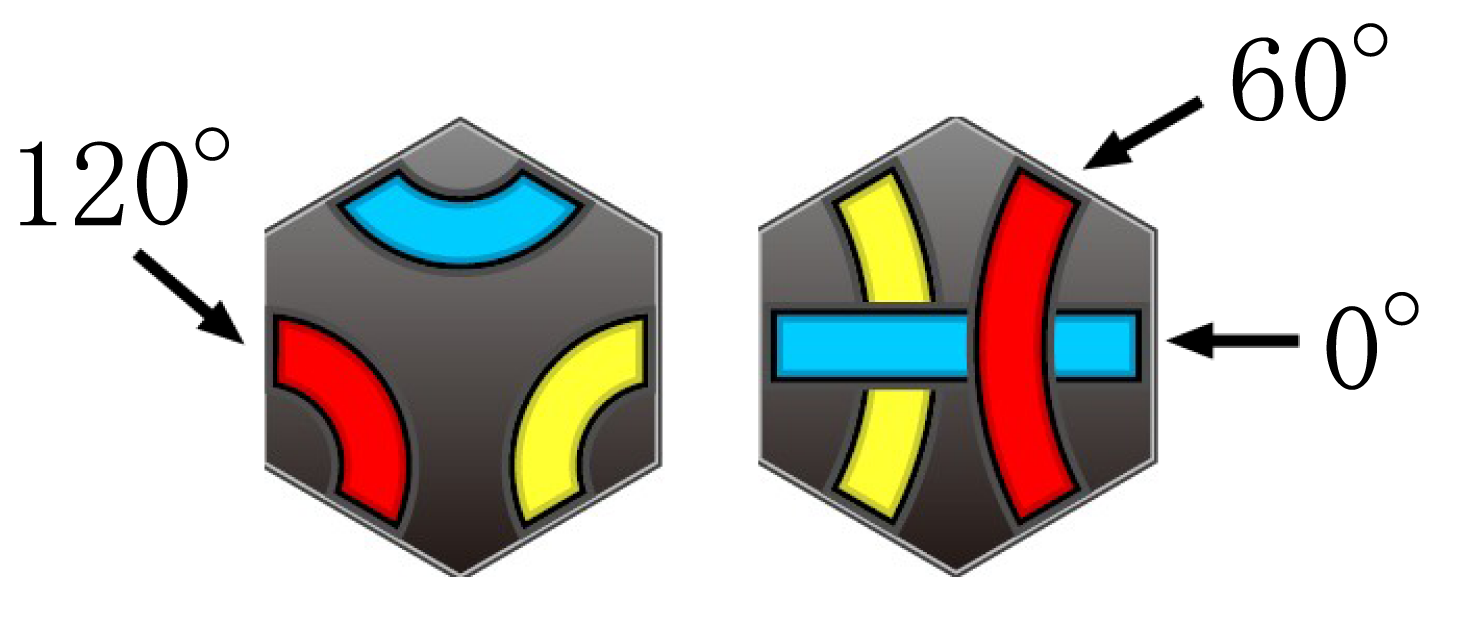}
 \end{center}
\vspace{-0.3cm}
  \caption{Three angles of lines.} 
\label{angles}
\end{minipage}
\begin{minipage}{.04\linewidth}
\mbox{}
\end{minipage}
\begin{minipage}{.345\linewidth}
 \begin{center}
  \includegraphics[height=16mm]{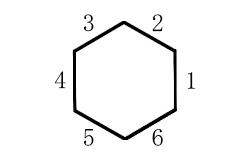}
 \end{center}
\vspace{-0.45cm}
  \caption{Numbers of edges of a place.}
 \label{edges}
\end{minipage}
\end{figure}

We call a tile with number $i$ on its back {\it tile $i$}. 
Remind that the designated color of a number is the one 
in which the number is written. 
We define an {\it angle} of a line 
as its central angle when we regard it as a circular arc of a circle, 
where we define the angle of a straight line to be $0^\circ$. 
Then an angle of a line 
is either one of $0^\circ $, $60^\circ $ or $120^\circ $ 
(Fig.~\ref{angles}). 
For example, tile 2 is said to have a red line of an angle $120^\circ $, 
$0^\circ $ blue line and $120^\circ $ yellow line 
(Fig.~\ref{tile_surfaces} (a)).

\begin{figure}[htb]
 \begin{center}
  \includegraphics[height=15.5mm]{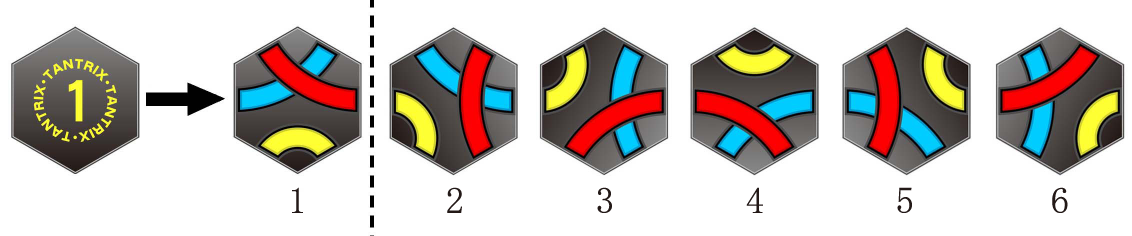}
 \end{center}
\vspace{-0.3cm}
  \caption{Orientations of a tile.} 
 \label{orientations}
\end{figure}

To distinguish places and orientations of tiles to be placed, 
we assign numbers to them. 
For places of an input board of size $m$, 
we give numbers from $1$ to $m$ as explained above 
(Fig.~\ref{spiral_board}). 
We also give numbers from 1 to 6 to the edges of each place 
couter-clockwise as shown in Fig.~\ref{edges}. 
The {\it orientations} of a tile to be placed take values from 1 to 6. 
We define it in the following way (Fig.~\ref{orientations}): 
place a tile showing its back and its number in the upright position; 
flip it horizontally, 
and it is orientation~1; 
every time we rotate it by $60^\circ $ clockwise, 
its orientations will be $2,3,4,5$ and $6$, respectively. 
Fig.~\ref{tile_surfaces} (a) shows each tile in its orientation~1. 

Two places are {\it adjacent} if they share an edge. 
Let $a(j, \ell )$ be a function 
that returns the number of the place to which place $j$ is adjacent 
with its edge $\ell$, 
and returns 0 if such a place does not exist (out of a board). 
For a board shown in Fig.~\ref{spiral_board} (a), for example, 
$a(4,6)=12$, $a(8,5)=0$, and so on. 
For simplicity, once a tile is (or supposed to be) placed 
on a certain place with a certain orientation, 
we allow to identify the tile with that place. 
That is, we can say: adjacent tiles; 
we can alternatively say 
that the color of a line of an edge of a place 
instead of the color of a line (of a placed tile) 
appearing on the corresponding edge 
(of the place where the tile is placed in a certain orientation), 
adjacent tiles, and so on. 
We say that a place is {\it empty} if no tile is placed on it. 

\section{An Integer Programming Formulation}
\label{formulation}

We solve this problem {\sc TANTRIX} 
by utilizing a mathematical programming solver 
after formulating it as an integer program. 
We first introduce its variables, 
and explain constraints and an objective function. 
Then we show preliminary experimental results 
of solving this formulated IP.

\subsection{Variables}

Remind here that $n$ is a challenge number, 
$m$ $(>n)$ is the size of a prepared board 
and that we use 10 sorts of tiles numbered from 1 to 10 on their backs. 
When the challenge number is $n$, 
the number $N$ of sorts of tiles to be used is $N=\min\{n,10\}$ $(\le 10)$, 
which implies that $N=10$ if $n\ge 10$. 
We assume that we try Tantrix of challenge number 
no less than 10 and that $N$ is always 10. 
Then we set the following three integral variables for our IP formulation. 

First, for each tile $i$ $(1\le i\le 10)$, 
for each place $j$ $(1\le j\le m)$ 
and for each orientation $k$ $(1\le k\le 6)$ of tiles, 
let a $0$-$1$ variable $x_{ijk}$ be 
\begin{eqnarray*}
x_{ijk}=
\left\{
\begin{array}{ll}
1, & \mbox{tile $i$ is placed on place $j$ with orientation $k$},\\
0, & \mbox{otherwise (i.e., tile $i$ is not placed on place $j$ with orientation $k$).} 
\end{array}
\right.
\end{eqnarray*}
Next for each place $j$ $(1\le j\le m)$ and its edge $\ell$ 
$(1\le \ell \le 6)$, 
let a variable $y_{j\ell }$ that expresses the color 
of its corresponding line (of a tile) be 
\begin{eqnarray*}
y_{j\ell}=
\left\{
\begin{array}{ll}
 0, & \mbox{for place $j$ and its edge $\ell$, 
there is no color}, \\ 
 1, & \mbox{for place $j$, the color of the line 
corresponding to edge $\ell$ } \\ 
    & \mbox{\ \ \ \ is neither the designated color nor color 2}, \\
 2, & \mbox{for place $j$, the color of the line 
corresponding to edge $\ell$ } \\ 
    & \mbox{\ \ \ \ is neither the designated color nor color 1}, \\
 3, & \mbox{for place $j$, the color of the line 
corresponding to edge $\ell$ } \\ 
    & \mbox{\ \ \ \ is the designated color.} 
\end{array}
\right.
\end{eqnarray*}
Here we use values 1, 2 and 3 to indicate three colors of lines, 
where 3 is intended to denote the designated color, 
and value 0 is prepared for empty places 
(thus no color is defined for their edges). 
Notice here that if we define a function $c(i, k, \ell )$ 
that returns the color of the line corresponding to edge $\ell$ 
when tile $i$ is placed in orientation $k$ 
(we can know it according to the top of tiles), 
the variable $y_{j\ell }$ is represented by using $x_{ijk}$ and $c(i,k,\ell)$ 
as follows: 
\[ 
\textstyle
y_{j\ell} = \sum _{i=1}^{n} \sum _{k=1}^{6} c(i, k, \ell) x_{ijk} 
\ \ \  (j=1, 2, \dots , m;~\ell=1, \dots , 6) .
\]

To describe some constraints succinctly, 
we introduce for convenience a 0-1 variable $u_{jj'}$ 
for each pair of adjacent places $j$ and $j'$: 
\begin{eqnarray*}
u_{jj'}=
\left\{
\begin{array}{ll}
 1, & \mbox{a tile is placed on either one of the places $j$ and $j'$} ,\\
 0, & \mbox{otherwise (i.e., tiles are placed both or neither 
on places $j$ and $j'$)}.
\end{array}
\right.
\end{eqnarray*}
That is, $u_{jj'}$ is simply the exclusive OR of two (0-1) variables 
$\sum_{i=1}^{10}\sum_{k=1}^{6} x_{ijk}$ and 
$\sum_{i=1}^{10}\sum_{k=1}^{6} x_{ij'k}$ 
which imply if a tile is placed on place $j$ and $j'$, respectively. 
The possible combinations of values of these two variables and $u_{jj'}$ are 
$(\sum_{i=1}^{10}\sum_{k=1}^{6} x_{ijk}$, 
$\sum_{i=1}^{10}\sum_{k=1}^{6}x_{ij'k}$, $u_{jj'})$
$\in\{(0, 0, 0)$, $(1, 1, 0)$, $(1, 0, 1)$, $(0, 1, 1)\}$, 
and we can confirm that these conditions are expressed 
only by using variable $x_{ijk}$ as in the following system 
of four inequalities: 
$u_{jj'} \leq - \sum_{i=1}^{10} \sum_{k=1}^{6} x_{ijk} - \sum_{i=1}^{10} \sum_{k=1}^{6} x_{ij'k} + 2$, 
$u_{jj'} \leq \sum_{i=1}^{10} \sum_{k=1}^{6} x_{ijk} + \sum_{i=1}^{10} \sum_{k=1}^{6} x_{ij'k}$, 
$ u_{jj'} \geq \sum_{i=1}^{10} \sum_{k=1}^{6} x_{ijk} - \sum_{i=1}^{10} \sum_{k=1}^{6} x_{ij'k}$ and 
$u_{jj'} \geq - \sum_{i=1}^{10} \sum_{k=1}^{6} x_{ijk} + 
 \sum_{i=1}^{10} \sum_{k=1}^{6} x_{ij'k}$.

\subsection{Constraints and an Objective Function}

We now describe the (necessary) conditions 
that have to be satisfied by Tantrix solutions 
as constraints of an integer program. 
Based on the rules of Tantrix, 
we introduce the following five constraints. 

\begin{quote}
{\bf Constraint 1} (C1). 
At most one tile is placed on each place. 
\end{quote}

\noindent
This is required to hold for any place on a board 
that two or more tiles cannot be placed, 
and it is represented by the following formula: 
\[
\mbox{C1:} \quad \textstyle\sum _{i=1}^{10}\sum _{k=1}^{6} x_{ijk}\le 1 
\quad (j=1, 2, \dots ,m). 
\]

\begin{quote}
{\bf Constraint 2} (C2). 
The number of places each on which a tile is placed 
equals the challenge number. 
\end{quote}

\noindent
This implies that we use exactly the same number of places 
as the challenge number, and is formulated as follows: 
\[
\mbox{C2:} \quad \textstyle
\sum _{i=1}^{10}\sum_{j=1}^{m}\sum _{k=1}^{6} x_{ijk} = n. 
\]

\begin{quote}
{\bf Constraint 3} (C3). 
Each tile is used the number of times defined by the challenge number. 
\end{quote}

\noindent
Once we determine the challenge number to be $n$, 
the number of tiles $i$ used for it is described by the following formula: 
\[
\mbox{C3:} \quad \textstyle\sum_{j=1}^{m}\sum _{k=1}^{6} x_{ijk}
=\left\lceil\frac{n+1-i}{10}\right\rceil \quad (i=1,\dots ,10). 
\]

\begin{quote}
{\bf Constraint 4} (C4). 
The color of a line of an edge that is adjacent to no other tile 
is not the designated color. \\
{\bf Constraint 5} (C5). 
The colors of lines whose corresponding edges are touching each other 
have to match. 
\end{quote}

\noindent
These constraints tell about two adjacent places $j$ and $j'$ 
via their edges $\ell$ and $\ell'$, respectively. 
For C4, if exactly one tile is placed on either of two places $j$ and $j'$, 
then either one of $y_{j\ell}$ or $y_{j'\ell'}$ is 0 
and the other is 1 or 2, 
i.e., if $u_{jj'}=1$ then $1\leq \vert y_{j\ell}-y_{j'\ell'}\vert \leq 2$. 
Similarly for C5, 
$y_{j\ell}=y_{j'\ell'}$ $=1,2 \mbox{ or } 3$ 
if tiles are placed both on $j$ and $j'$ 
and $y_{j\ell}=y_{j'\ell'}=0$ otherwise, 
i.e., if $u_{jj'}=0$ then $0\leq \vert y_{j\ell}-y_{j'\ell'}\vert \leq 0$ 
in any case. 
By paying attention to that $y_{j\ell}\neq y_{j'\ell'}$ if $u_{jj'}=1$, 
we can formulate these two constraints together as follows: 
\[
\mbox{C4+C5:}\quad -2u_{jj'}\leq y_{j\ell }-y_{j'\ell'} 
\leq 2u_{jj'}\quad (a(j, \ell )=j';a(j', \ell')=j).
\]

Remark here that by C4 and C5, any line of the designated color 
must be a part of a loop, 
that is, they constitute necessary (but not sufficient) conditions 
for that all the lines of the designated color form loops, 
but that they do not imply sufficient conditions of Tantrix 
solutions in the sense that they may have holes 
or that a loop of the designated color may not be unique. 
We call such a loop of the designated color consists of less 
than $n$ tiles a {\it subloop}. 
We do not expect to have complete necessary and sufficient conditions 
in our elementary formulation, 
and we will deal with these issues 
(ad-hoc or try and error, in some sense) later in subsequent sections. 

In our approach, 
since a solution is formulated as constraints of an integer program, 
it suffices to find one of its feasible solutions. 
In such a case, an objective function is enough to be virtual, 
and therefore, we set it as $x_{1,1,1}\rightarrow \min\!.$, 
for a descriptive purpose.

\subsection{Elementary Experiments and the Results}

We solve the integer program formulated so far 
by a mathematical programming solver, 
to obtain a solution and to examine the solution time. 
We use IBM ILOG CPLEX 12.2 \cite{cplex} as a solver, 
which is installed on a single PC of Intel Pentium Dual E2200 processor 
(2.2GHz) with 1GB RAM. 
Since the computational time spent for solving a fixed formulation 
by a solver on a same computer is always the same in principle, 
we adopt the one by a single computation throughout our experiments. 

We performed experiments for challenge numbers 10, 15 and 20 
with appropriate board sizes. 
(We sometimes denote the input challenge number $n$ 
and its board size $m$ for experiments by $(n,m)$, for short.) 
Table~\ref{C1-C5} shows its computational time, the number of holes 
and the number of loops in each solution. 
We see that we can solve our formulation within 30 seconds 
up to challenge number 20. 
As we mentioned, constraints C1--C5 cannot eliminate 
holes and subloops in their solutions, and in fact, 
for challenge numbers 10, 15 and 20 all the solutions include holes 
and one solution includes a subloop as well (Fig.~\ref{15-19/C1-C5}). 
These results imply that it is not so easy even to obtain (valid) 
Tantrix solutions only by these elementary constraints. 
Therefore, we concentrate on deriving such formulations 
that can produce valid solutions 
rather than obtain them in less computational time. 

\begin{table}[htb]
\begin{minipage}{0.63\hsize}
\begin{center}
\caption{Computational results of solving IP formulations of {\sc TANTRIX} 
using constraints C1--C5: computational time (sec.), 
the number of holes and loops for each challenge number $n$ 
with its board size $m$.}
\label{C1-C5}
\small 
\begin{tabular}{|c|c||r|c|c|} \hline
$n$ & $m$ & \multicolumn{1}{c|}{time} & \#holes & \#loops \\ \hline \hline
10 & 12 & 0.27 & 1 & 1 \\ \cline{2-5}
   & 19 & 1.64 & 1 & 1 \\ \hline
15 & 19 & 0.84 & 1 & 2 \\ \cline{2-5}
   & 27 & 9.64 & 1 & 1 \\ \hline
20 & 27 & 27.41 & 1 & 1 \\ \cline{2-5}
   & 37 & 12.36 & 1 & 1\\ \hline
\end{tabular}
\vspace*{-0.3cm}
\end{center}
\end{minipage}
\begin{minipage}{0.02\hsize}\mbox{}
\end{minipage}
\makeatletter
\def\@captype{figure}
\makeatother
\begin{minipage}{0.33\hsize}
\begin{center}
\vspace*{0.3cm}
  \includegraphics[height=34mm]{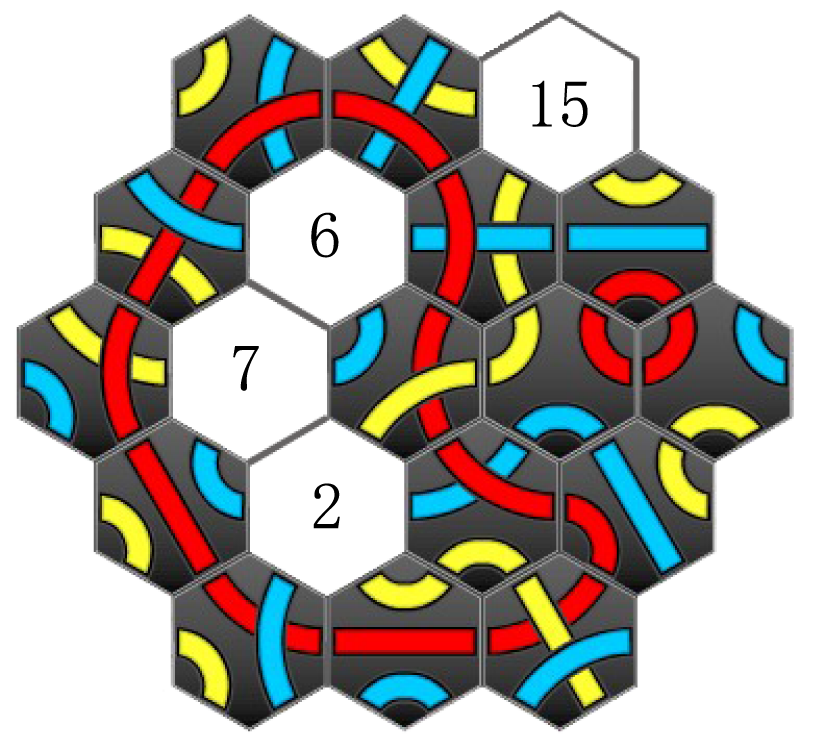}
\end{center}
\caption{The solution for $(15,19)$; it has a hole and two subloops. }
\label{15-19/C1-C5}
\vspace*{-0.3cm}
\end{minipage}
\end{table}

\section{Improvement of Formulations for Valid Tantrix Solutions}
\label{improvement}

The IP formulations which we proposed 
in the previous section 
cannot always lead Tantrix solutions even 
for relatively small challenge numbers 
mainly due to holes and subloops. 
To deal with these issues and to obtain valid solutions, 
in this section, we consider additional constraints 
for (i) arranging solution shapes to be `round' (and thus holes may disappear) 
and (ii) eliminating subloops in advance.

\subsection{Arranging Solution Shapes to be Round and Holeless}

Rules of Tantrix implicitly require the solution shapes to be round 
and this automatically implies that there can hardly be holes in solutions. 
Hence, to avoid holes, 
we propose two ideas for keeping solution shapes to be round. 

\subsubsection{Restrict the number of adjacent tiles}

A hole is a set of adjacent empty places surrounded by 6 or more tiles. 
By utilizing this observation, 
we first consider the following three constraints. 

\begin{quote}
{\bf Constraint 6a} (C6a). 
At most 5 tiles are place on adjacent places to an empty place. \\
{\bf Constraint 6b} (C6b). 
At most 4 tiles are place on adjacent places to an empty place. \\
{\bf Constraint 6c} (C6c). 
At most 3 tiles are place on adjacent places to an empty place. 
\end{quote}

\noindent
These are expressed in the following formulas: 
\[
\begin{array}{l}
\mbox{C6a:} \quad \sum_{\ell=1}^{6}\sum _{i=1}^{10}\sum _{k=1}^{6} 
x_{i, a(j,\ell), k} \leq  \sum _{i=1}^{n} \sum _{k=1}^{6} x_{ijk} + 5 
\quad (j=1,\dots ,m), \\
\mbox{C6b:} \quad \sum_{\ell=1}^{6}\sum _{i=1}^{10}\sum _{k=1}^{6} 
x_{i, a(j,\ell), k} \leq  2\sum _{i=1}^{n}\sum _{k=1}^{6} x_{ijk} + 4 
\quad (j=1,\dots ,m), \\
\mbox{C6c:} \quad \sum_{\ell=1}^{6}\sum _{i=1}^{10}\sum _{k=1}^{6} 
x_{i, a(j,\ell), k} \leq  3\sum _{i=1}^{n}\sum _{k=1}^{6} x_{ijk} + 3 
\quad (j=1,\dots ,m). 
\end{array}
\]

\noindent 
The lefthand side of the formulas counts the number of tiles 
placed on adjacent places to a place $j$, 
and each righthand side becomes 5 (4, 3, respectively) 
if a tile is not placed on place $j$, and 6 otherwise. 
Thus the inequalities directly imply the constraints.

\begin{figure}[htb]
 \begin{center}
  \includegraphics[height=42.5mm]{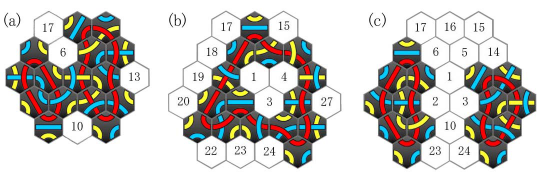}
 \end{center}
\caption{The solution for (a) $(15,19)$ with C6a, 
(b) $(15,27)$ with C6b and (c) $(15,27)$ with C6c.}
\label{solC1-C5+C6}
\end{figure}

Since C6c implies C6b, and C6b implies C6a 
(i.e., C6c $\Rightarrow$ C6b $\Rightarrow$ C6a), 
we use either one of them together with C1--C5, 
and adopt an effective one (in a way, try and error). 
We show in Table~\ref{C1-C5+C6} computational results 
of solving formulations 
using C1--C5 with adding each one of C6a, C6b or C6c, 
and Fig.~\ref{solC1-C5+C6} shows some of their solutions. 
As a result, we see that we can obtain Tantrix solutions 
only for challenge numbers 10 and 15, 
and cannot yet for challenge number 20. 
In these solutions, not only there are subloops but the shapes seem irregular 
(not round (Fig.~\ref{solC1-C5+C6} (a)) 
or a hole exists (Fig.~\ref{solC1-C5+C6} (b))), 
and even worse, the shape becomes disconnected unexpectedly 
(Fig.~\ref{solC1-C5+C6} (c)) although it does not have a hole. 
According to these observations, we see that C6's are not 
effective enough to avoid holes or subloops.

\begin{table}[htb]
\begin{center}
\caption{Computational results of solving IP formulations 
using C1--C5 with C6a, C6b or C6c 
for challenge numbers 10, 15 and 20. 
Here $*$ implies that the solution is disconnected.}
\label{C1-C5+C6}
\small
\begin{tabular}{|c|c||r|c|c|r|c|c|r|c|c|} \hline
\multicolumn{2}{|c||}{} & \multicolumn{3}{c|}{C6a} & \multicolumn{3}{c|}{C6b} & \multicolumn{3}{c|}{C6c} \\ \hline
$n$ & $m$ & \multicolumn{1}{c|}{time} & \#holes & \#loops & \multicolumn{1}{c|}{time} & \#holes & \#loops & \multicolumn{1}{c|}{time} & \#holes & \#loops \\ \hline\hline
10 & 12 & 0.27 & 0 & 1 & \multicolumn{1}{c|}{------} & ------ & ------ & \multicolumn{1}{c|}{------} & ------ & ------ \\ \cline{2-11}
   & 19 & 2.19 & 0 & 1 & \multicolumn{1}{c|}{------} & ------ & ------ & \multicolumn{1}{c|}{------} & ------ & ------ \\ \hline
15 & 19 & 17.72 & 0 & 2 & 16.14 & 0 & 1 & 5.27 & 0 & 2 \\ \cline{2-11}
   & 27 & 17.24 & 1 & 2 & 4.51 & 1 & 1 & 63.89 & $0^*$ & 2 \\ \hline
20 & 27 & 5.69 & 1 & 1 & 35.83 & 0 & 2 & 10.47 & 0 & 2 \\ \cline{2-11}
   & 37 & 12.59 & 1 & 1 & 98.67 & 0 & 2 & 509.11 & $0^*$ & 3 \\ \hline
\end{tabular}
\end{center}
\end{table}

\subsubsection{Utilize the objective function}

The second idea for arranging solution shapes is to change 
the objective function so that it forces solution shapes to be round. 
For this purpose, we define a weight $w(j)$ for each place $j$ on a board; 
that is, if place $j$ is on $r$-th round from its center 
(which is defined to be 0-th round), then $w(j)=-r$. 
For a board of type A (Fig.~\ref{spiral_board} (a)) 
and for its place 8, for example, 
since it is on 2nd round, $w(8)=-2$. 
Then we set a new objective function to be 
\[
\textstyle
\sum_{j=1}^{m}\sum_{i=1}^{10}\sum_{k=1}^{6}w(j)x_{ijk}\rightarrow \mbox{max}. 
\]
(instead of $x_{1,1,1}\rightarrow$ min.). 
This is intended for solutions to use inner places 
more likely than outer places, 
and therefore we expect that solution shapes become round 
and are unlikely to have holes. 

Table~\ref{C1-C5+weight} shows computational results of solving formulations 
using this objective function with constraints C1--C5, 
and Fig.~\ref{solC1-C5+weight} shows some of these solutions. 
We can see that their shapes seem to become rather round compared to those 
obtained by the previous idea, however, 
we still have holes in some cases. 

\begin{table}[htb]
\begin{minipage}{0.33\hsize}
\begin{center}
\caption{Computational results of solving IP formulations 
using C1--C5 with a weighted objective function 
for challenge numbers 15 and 20.}
\vspace{3mm}
\label{C1-C5+weight}
\small \begin{tabular}{|c|c||r|c|c|} \hline
$n$& $m$ & \multicolumn{1}{c|}{time} & \#holes & \#loops \\ \hline\hline
 15& 19& 27.44 & 0 & 1\\ \cline{2-5}
  & 27& 6.70&1&1\\ \hline
  20& 27&40.17 &1&1\\ \cline{2-5}
 & 37& 13.20&0&3\\ \hline
\end{tabular}
\end{center}
\end{minipage}
\begin{minipage}{0.02\hsize}\mbox{}
\end{minipage}
\makeatletter
\def\@captype{figure}
\makeatother
\begin{minipage}{0.63\hsize}
 \begin{center}
  \includegraphics[height=49.5mm]{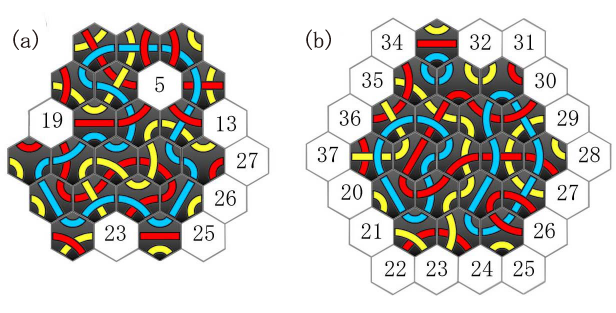}
 \end{center}
\caption{
The solutions for (a) $(20,27)$ and (b) $(20,37)$ 
with a weighted objective function.}
\label{solC1-C5+weight}
\end{minipage}
\end{table}

\subsubsection{Incorporating two ideas simultaneously}

As we observed, 
the two proposed ideas for arranging solution shapes 
are not always effective when each of them is used alone. 
Therefore, we try to implement these simultaneously; 
that is, we adopt the weighted objective function 
and also add either one of constraints C6a, C6b or C6c. 
We show computational results for challenge numbers 15 and 20 
in Table~\ref{C1-C6+weight}. 
Then the effect of combining two ideas seems remarkable 
as holes disappear completely from all solutions, 
and 
as a result, we can have at least one Tantrix solution 
for challenge numbers up to 20. 

\begin{figure}[htb]
 \begin{center}
  \includegraphics[height=48mm]{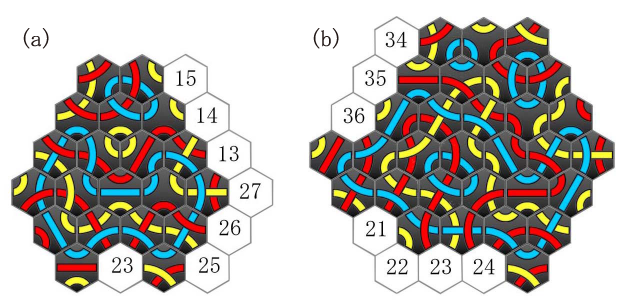}
 \end{center}
\caption{
The solutions for (a) $(20,27)$ with C6a 
and (b) $(30,37)$ with C6c.}
\label{solC1-C6+weight}
\end{figure}

Then we try some larger challenge numbers, 25 and 30, under this formulation, 
and we show those computational results also 
in Table~\ref{C1-C6+weight} and Fig.~\ref{solC1-C6+weight}. 
The solution shapes seem to be well arranged 
and there are no holes in any solutions at all, 
which again confirms the effect of combining two ideas. 
However, we cannot have Tantrix solutions for challenge number 30 
due to subloops. 
We approach this issue in the next subsection.

\begin{table}[htb]
\begin{center}
\caption{
Computational results of solving IP formulations using C1--C6 
with a weighted objective function for challenge numbers 15, 20, 25 and 30.}
\label{C1-C6+weight}
\small \begin{tabular}{|c|c||r|c|c|r|c|c|r|c|c|} \hline
\multicolumn{2}{|c||}{}& \multicolumn{3}{c|}{C6a}& \multicolumn{3}{c|}{C6b}& \multicolumn{3}{c|}{C6c}\\ \hline
$n$& $m$ & \multicolumn{1}{c|}{time} & \#holes & \#loops & \multicolumn{1}{c|}{time} & \#holes & \#loops & \multicolumn{1}{c|}{time} & \#holes & \#loops \\ \hline\hline
  15& 19& 3.28 &0&2 & 29.58 &0& 1& 5.92 &0&2 \\ \cline{2-11}
 & 27& 36.75 &0& 1& 12.53 &0& 1& 29.84&0&  2\\ \hline
 20& 27& 5.73 & 0&2& 7.34  &0& 2& 10.13 &0&1 \\ \cline{2-11}
 & 37& 405.61 &0& 2& 26.13 &0&1 & 18.70 &0& 1\\ \hline\hline
 25& 37& 243.33 &0&1 & 79.16 &0& 1& 525.28 &0& 1\\ \hline
 30& 37& 74.39&0& 3& 415.72 &0& 4& 26.19 &0& 3\\ \cline{2-11}
 & 48& 72.55&0& 4& 624.52&0& 2& 21.58 &0& 2\\ \hline
\end{tabular}
\end{center}
\end{table}

\subsection{Eliminate Short Subloops in Advance}

So far, 
we achieved to arrange solution shapes 
to be round and holeless to some extent. 
On the other hand, subloops can appear in solutions 
since we do not express the uniqueness of a loop as constraints 
of our formulations. 
It is, of course, not impossible to introduce constraints 
that eliminates all the probable subloops 
as introduced and well studied for TSP formulations 
\cite{ABCC06,CCPS97,P03,W99}. 
However, it may require exponential number of constraints, 
which may also cause a long computational time. 
Furthermore, since a solution shape is not determined to be unique 
due to the surplus of places 
prepared for a challenge number to guarantee the freedom of solution shapes, 
the cut which has to be crossed by a loop in a solution 
is no longer a cut for (a shape of) another solution. 
This makes it not so easy to take an approach 
that we add constraints for eliminating subloops 
every time they appear and {\rm re-solve} it 
to obtain a unique loop of the designated color \cite{KU11}. 

Now observing solutions that have subloops, 
we can see those consists of 3 or 4 tiles. 
Therefore, we attempt to 
eliminate these types of short subloops. 
More specifically, we eliminate subloops consists of 3, 4 or 5 tiles 
and describe it as the following constraints: 

\begin{quote}
{\bf Constraint 7} (C7). 
There are no loops consists of 3 tiles. 

{\bf Constraint 8} (C8). 
There are no loops consists of 4 tiles. 

{\bf Constraint 9} (C9). 
There are no loops consists of 5 tiles. 
\end{quote}

\begin{figure}[htb]
 \begin{center}
  \includegraphics[height=48mm]{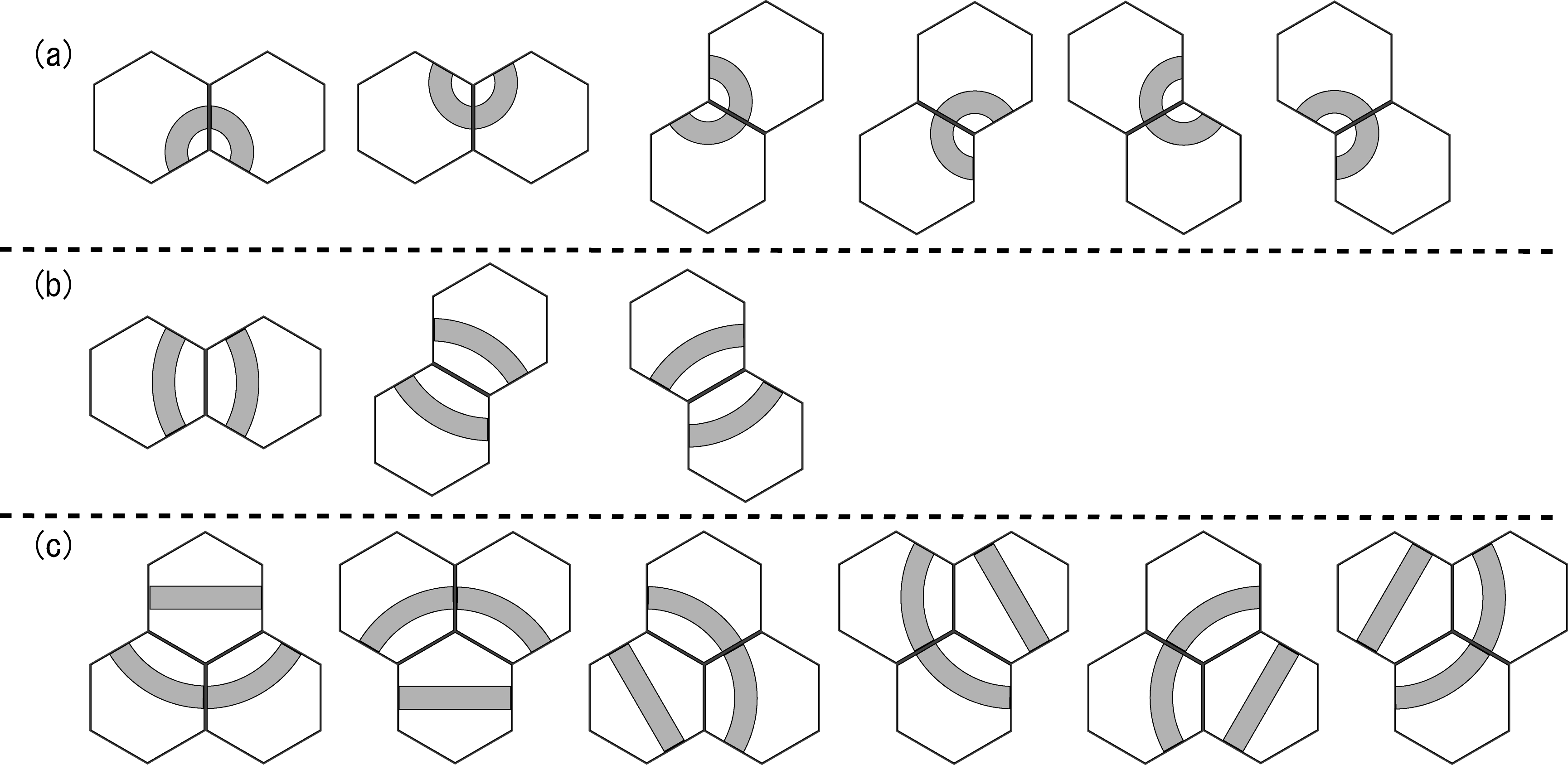}
 \end{center}
  \caption{
Forbidden arrangements of lines of the designated color 
(for challenge number greater than 5). 
These will force subloops consist of 
(a) 3, (b) 4 and (c) 5 tiles, respectively.} 
 \label{forbidden_arrangements}
\end{figure}

We achieve them by embedding these constraints 
for every possible places on a board in advance. 
It is easy to see that C7 is equivalent to that none of the arrangements 
of lines of the designated color 
shown in Fig.~\ref{forbidden_arrangements} (a) appears 
on any two adjacent places on a board. 
Therefore, this constraint is realized by forbidding these types of tiles 
(having $120^\circ$ lines of the designated color) 
to become one of those arrangements.

Suppose now that the designated color is red. 
We can see that tiles 2 and 3 have $120^\circ$ red lines 
(Fig.~\ref{tile_surfaces}). 
To forbid these two tiles to be placed in one of those arrangements 
simultaneously, 
we force the sum of $x_{ijk}$ corresponding to being these arrangements 
not to be greater than 1. 
Therefore, in this concrete case, the formulations become as follows. 
We can consider similarly in the case 
that the designated color is blue or yellow: 
\begin{eqnarray*}
\left\{
\begin{array}{l}
  x_{2,j,2} + x_{3,j,4} + x_{2,a(j, 1),6} + x_{3,a(j, 1),2} \leq 1 \quad (j=1,\dots, m),\\
   x_{2,j,3} + x_{3,j,5} + x_{2,a(j, 1),5} + x_{3,a(j, 1),1}\leq 1 \quad (j=1,\dots, m ),\\
   x_{2,j,1} + x_{3,j,3} + x_{2,a(j, 2),5} + x_{3,a(j, 2),1} \leq 1\quad (j=1,\dots, m),\\
   x_{2,j,2} + x_{3,j,4} + x_{2,a(j, 2),4} + x_{3,a(j, 2),6} \leq 1\quad (j=1,\dots, m ),\\
   x_{2,j,6} + x_{3,j,2} + x_{2,a(j, 3),4} + x_{3,a(j, 3),6} \leq 1\quad (j=1,\dots, m ),\\
   x_{2,j,1} + x_{3,j,3} + x_{2,a(j, 3),3} + x_{3,a(j, 3),5} \leq 1\quad (j=1,\dots, m ). 
\end{array}
\right.
\end{eqnarray*}
%
\nop{
For constraints C8 and C9, 
we can eliminate these types of subloops in a quite similar way 
according to the observations shown in Fig.~\ref{forbidden_arrangements} 
(b) and (c). 
(See Appendix~\ref{4+5subloops} for details.) 
}

For constraint C8, we can deal with it in a similar way to C7. 
In fact, C8 is equivalent to that none of the three arrangements of lines 
of the designated color in Fig.~\ref{forbidden_arrangements} (b) appears 
on any two adjacent places on a board. 
Therefore, it is realized by forbidding these types of tiles 
(having $60^\circ$ lines of the designated color) 
to become one of those arrangements. 
It is also the case that we can represent this constraint 
by the corresponding variables $x_{ijk}$, 
and in the case of the designated color is red, for example, 
it is formulated as follows 
(similarly for blue and yellow): 
\begin{eqnarray*}
\left\{
\begin{array}{l}
 x_{1,j,2} + x_{4,j,1} + x_{6,j,4} + x_{7,j,2} + x_{8,j,3} + x_{10,j,3} + x_{1,a(j, 1),5}+ x_{4,a(j, 1),4}\\ 
\ \ \ \  + x_{6,a(j, 1),1} 
 + x_{7,a(j, 1),5} +  x_{8,a(j, 1),6} + x_{10,a(j, 1),6}\leq 1 \quad (j=1,\dots, m),\\
 x_{1,j,1} + x_{4,j,6} + x_{6,j,3} + x_{7,j,1} + x_{8,j,2} + x_{10,j,2} + x_{1,a(j, 2),4} + x_{4,a(j, 2),3}\\ 
\ \ \ \ + x_{6,a(j, 2),6}
 + x_{7,a(j, 2),4} + x_{8,a(j, 2),5} + x_{10,a(j, 2),5} \leq 1\quad (j=1,\dots, m),\\
 x_{1,j,6} + x_{4,j,5} + x_{6,j,2} + x_{7,j,6} + x_{8,j,1} + x_{10,j,1} + x_{1,a(j, 3),3} + x_{4,a(j, 3),2}\\
\ \ \ \   + x_{6,a(j, 3),5} + x_{7,a(j, 3),3} +  x_{8,a(j, 3),4} + x_{10,a(j, 3),4}\leq 1 \quad( j=1,\dots, m). 
\end{array}
\right.
\end{eqnarray*}

For C9, it is equivalent to that for lines of the designated color 
none of the six 
arrangements in Fig.~\ref{forbidden_arrangements} (c) appears 
on every mutually adjacent three places on a board. 
Suppose the designated color is red, 
then tiles 5 and 9 have $0^\circ$ red lines 
and tiles 1, 3, 6, 7, 8 and 10 have $60^\circ$ red lines. 
Therefore, we can express by using variables $x_{ijk}$ 
corresponding to these tiles 
that they do not form none of the arrangements, 
and in the case of the designated color is red, for example, 
it is formulated as follows 
(similarly for blue and yellow):

\vspace{-4mm}
\begin{eqnarray*}
\left\{
\begin{array}{l}
x_{5,j,2} + x_{5,j,5} + x_{9,j,2} + x_{9,j,5} + x_{1,a(j, 1),5} + x_{4,a(j, 1),4} + x_{6,a(j, 1),1}+ x_{7,a(j, 1),5}\\
\ \ \ \  + x_{8,a(j, 1),6}+ x_{10,a(j, 1),6}+ x_{1,a(j, 2),4} + x_{4,a(j, 2),3} + x_{6,a(j, 2),6}+ x_{7,a(j, 2),4}\\
\ \ \ \  +x_{8,a(j, 2),5}+ x_{10,a(j, 2),5}\leq 2 \quad (j=1,\dots, m),\\

x_{1,j,1} + x_{4,j,6} + x_{6,j,3}+ x_{7,j,1} + x_{8,j,2} + x_{10,j,2} +x_{1,a(j, 1),6} + x_{4,a(j, 1),5}\\
\ \ \ \  + x_{6,a(j, 1),2}+ x_{7,a(j, 1),6} + x_{8,a(j, 1),1}+ x_{10,a(j, 1),1} +  x_{5,a(j, 2),1} + x_{5,a(j, 2),4}\\
\ \ \ \  + x_{9,a(j, 2),1} + x_{9,a(j, 2),4}\leq 2 \quad (j=1,\dots, m),\\
      
 x_{1,j,2} + x_{4,j,1} + x_{6,j,4}+ x_{7,j,2} + x_{8,j,3}+ x_{10,j,3}x_{5,a(j, 1),3} + x_{5,a(j, 1),3} \\
\ \ \ \  + x_{9,a(j, 1),3} + x_{9,a(j, 1),6}  + x_{1,a(j, 2),4}+ x_{4,a(j, 2),3} + x_{6,a(j, 2),6}+ x_{7,a(j, 2),4}\\
\ \ \ \  + x_{8,a(j, 2),5} + x_{10,a(j, 2),5}\leq 2 \quad (j=1,\dots, m),\\
      
 x_{5,j,1} + x_{5,j,4} + x_{9,j,1} + x_{9,j,4} + x_{1,a(j, 2),4} + x_{4,a(j, 2),3} + x_{6,a(j, 2),6}+ x_{7,a(j, 2),4}\\
\ \ \ \   + x_{8,a(j, 2),5}+ x_{10,a(j, 2),5} + x_{1,a(j, 3),3}  + x_{4,a(j, 3),2} + x_{6,a(j, 3),5}+ x_{7,a(j, 3),3}\\
\ \ \ \  + x_{8,a(j, 3),4} + x_{10,a(j, 3),4}\leq 2 \quad (j=1,\dots, m),\\

x_{1,j,6} + x_{4,j,5} + x_{6,j,2}+ x_{7,j,6} + x_{8,j,1} + x_{10,j,1} +x_{1,a(j, 2),5} + x_{4,a(j, 2),4}\\
\ \ \ \  + x_{6,a(j, 2),1}+ x_{7,a(j, 2),5} + x_{8,a(j, 2),6}+ x_{10,a(j, 2),6} +  x_{5,a(j, 3),3} + x_{5,a(j, 3),6}\\
\ \ \ \  + x_{9,a(j, 3),3} + x_{9,a(j, 3),6}\leq 2 \quad (j=1,\dots, m),\\

 x_{1,j,1} + x_{4,j,6} + x_{6,j,3}+ x_{7,j,1} + x_{8,j,2}+ x_{10,j,2}+ x_{5,a(j, 2),2} + x_{5,a(j, 2),5}\\
\ \ \ \ + x_{9,a(j, 2),2} + x_{9,a(j, 2),5} + x_{1,a(j, 3),2} + x_{4,a(j, 3),1} + x_{6,a(j, 3),4}+ x_{7,a(j, 3),2}\\
\ \ \ \  + x_{8,a(j, 3),3} + x_{10,a(j, 3),3}\leq 2 \quad (j=1,\dots, m).
\end{array}
\right.
\end{eqnarray*}
\vspace{1mm}

For challenge number 30, all of whose solutions have subloops, 
we show the results of solving formulations using C1--C9 
in Table~~\ref{C1-C9+weight}, 
where we omit the number of holes since all the solutions 
no longer have holes. 
We can see that 
we succeed to have Tantrix solutions for challenge number 30 
by eliminating short subloops.

\subsection{Further Challenge}

\begin{figure}[htbp]
 \begin{center}
  \includegraphics[height=66mm]{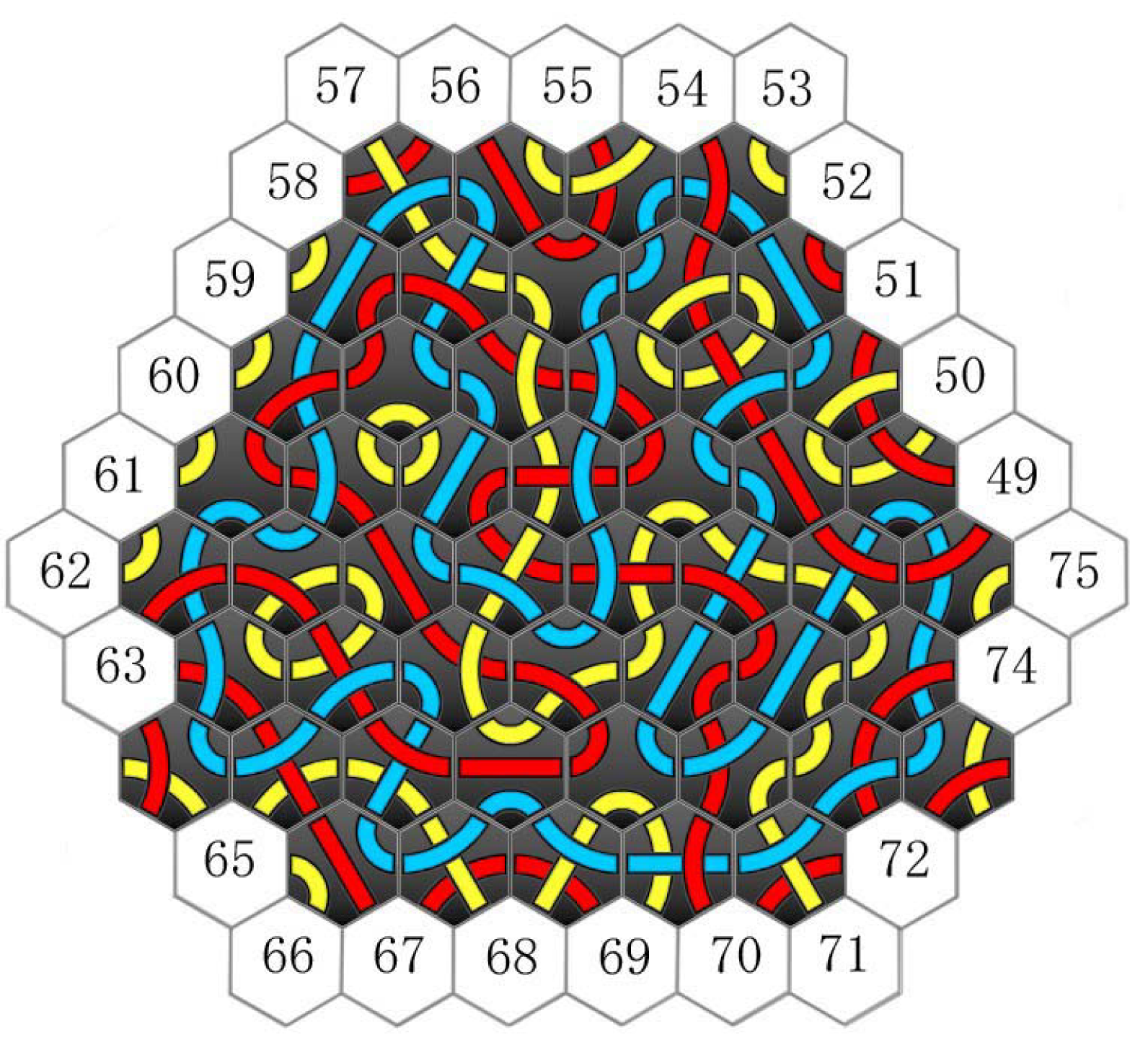}
 \end{center}
\caption{A Tantrix solution of challenge number 50 on a board of size 75, 
which is obtained by solving IP formulations.}
 \label{challenge50}
\end{figure}

To summarize all the ideas, observations and experiments presented so far, 
it gives good results that a formulation 
using constraints C1--C9 with weighted objective function is solved. 
Hence, under this formulation, we try to solve {\sc TANTRIX} 
by increasing its challenge numbers as 35, 40, 45 and 50. 
Table~\ref{C1-C9+weight} also shows these experimental results. 
As a consequence, the current best result (in the challenge number) is 50 
which is solved in total time 5549.20 seconds. 
We show the solution in Fig.~\ref{challenge50}. 

\begin{table}[htb]
\begin{center}
\caption{
Computational results of solving IP formulations using C1--C9 
with a weighted objective function 
for challenge numbers 30, 35, 40, 45 and 50.}
\label{C1-C9+weight}
\vspace{-0.2cm}
\small \begin{tabular}{|c|c||r|c|r|c|r|c|} \hline
\multicolumn{2}{|c||}{}& \multicolumn{2}{c|}{C6a}& \multicolumn{2}{c|}{C6b}& \multicolumn{2}{c|}{C6c}\\ \hline
$n$& $m$ & \multicolumn{1}{c|}{time} & \#loops & \multicolumn{1}{c|}{time} & \#loops & \multicolumn{1}{c|}{time} & \#loops \\ \hline \hline 
 30& 37&  89.63  & 1 &37.47 &1 &48.28 &1\\ \cline{2-8}
 & 48& 444.09  & 1 &1482.56&1& 4520.75 &1\\ \hline\hline
 35& 48& 25571.17& 1& 48.67&2& 917.42& 1\\ \hline
 40& 48& 1400.55& 2 &327.80& 1&665.02 &2\\ \hline
 45& 48& \multicolumn{1}{c|}{------} & ------&4698.13&3 & \multicolumn{1}{c|}{------} & ------\\ \cline{2-8}
 & 61& \multicolumn{1}{c|}{------} & ------&12747.89&1& \multicolumn{1}{c|}{------} & ------ \\ \hline
 50& 61& 408.09& 3&704.01&4 &2459.06& 2\\ \cline{2-8}
 & 75& \multicolumn{1}{c|}{------} & ------ &5549.20 &1& \multicolumn{1}{c|}{------} & ------ \\ \hline
\end{tabular}
\end{center}
\end{table}

\section{Conclusion}
\label{conclusion}

In this paper, we tried to solve Tantrix by a computer. 
The approach we adopted here is to formulate it as an IP 
and to solve it by a mathematical programming solver. 
We believe that this approach to solving puzzles 
is quite unique and entertaining as well. 
As a result, 
we could successfully solve it for challenge numbers up to 50, 
which is more or less larger than we expected. 
The results in this paper show 
that an approach using IP to solving puzzles appears promising 
and we believe that it may be valid for solving other puzzles. 
One of the important future work, of course, 
is to develop more effective formulations 
to solve Tantrix of larger challenge numbers in less computational time.

%



\newpage 
\appendix
\pagenumbering{Roman}
\markboth{APPENDIX}{APPENDIX}

\noindent{\large\bf APPENDIX}

\section{The Four-Tiles' Equation}
\label{4tiles_equation}

To be authorized as an official record, 
an arrangement must satisfy the following condition: 
in the arrangement let one of the three directions (axises) 
that has the most number $x$ of tiles be A, 
and the other two directions that cross A be B and C. 
Then there must be more tiles than 30\% of $x$ in more rows 
than 75\% of those in directions B and C, respectively. 
This condition is set to exclude the so-called ``4-tiles' equation'' 
discovered by C.~Fraser of England. 

\begin{figure}[htbp]
 \begin{center}
  \includegraphics[height=83mm]{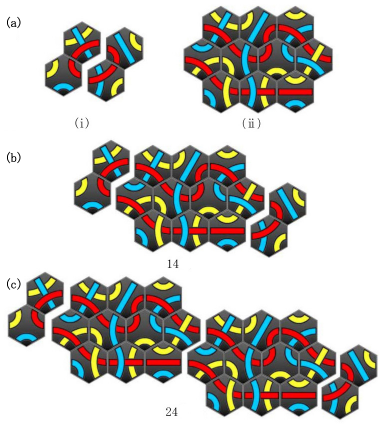}
 \end{center}
\caption{The 4-tiles' equation.}
 \label{4tiles}
\end{figure}

The 4-tiles' equation says that we can have a solution 
for an arbitrary large challenge number $10x+4$ $(x=0,1,\dots)$. 
In these cases the designated color is red. 
Fig.~\ref{4tiles} (a)(i) shows a red loop consists of four tiles 
numbered from 1 to 4, 
and (a)(ii) shows two red lines, which will be a part of a longer loop, 
consists of ten tiles numbered from 1 to 10. 
By combining these two parts, 
we will arrange infinitely large number of tiles into a single red loop, 
a solution for challenge numbers 14 and 24 in Fig.~\ref{4tiles} (b) and (c), 
respectively. 

This information was once posted in \cite{Tantrix-JP}, however, 
we can no longer find it on Jan. 1, 2012. 

\end{document}